\documentclass[twocolumn]{aastex631}

\usepackage{amsmath}
\usepackage{amssymb}
\usepackage{braket}
\usepackage{caption}
\usepackage{graphicx}
\usepackage{listings}
\usepackage{tikz}
\usepackage{verbatim}
\usepackage{xcolor}

\definecolor{keywordcolor}{rgb}{0.0, 0.0, 0.8}  % Blue for keywords
\definecolor{stringcolor}{rgb}{0.6, 0.2, 0.0}   % Brown for strings
\definecolor{commentcolor}{rgb}{0.0, 0.5, 0.0}  % Green for comments
\definecolor{backgroundcolor}{rgb}{0.95, 0.95, 0.95}  % Light gray background

\graphicspath{{./}{figures/}}

\begin{document}

\title{Not-Quite-Transcendental Functions For Logarithmic Interpolation of Tabulated Data}

\correspondingauthor{Jonah M.\ Miller}
\email{jonahm@lanl.gov}

\author[0000-0002-9447-1043
]{Peter C. Hammond}
\affiliation{Department of Physics, The Pennsylvania State University, University Park, PA 16802, USA}
\affiliation{Institute for Gravitation and the Cosmos, The Pennsylvania State University, University Park, PA 16802, USA}
\affiliation{Department of Physics and Astronomy, University of New Hampshire, Durham, NH 03824, USA}

\author[0000-0001-5705-1712]{Jacob M. Fields}
\affiliation{Department of Physics, The Pennsylvania State University, University Park, PA 16802, USA}
\affiliation{Institute for Gravitation and the Cosmos, The Pennsylvania State University, University Park, PA 16802, USA}

\author[0000-0001-6432-7860]{Jonah M. Miller}
\affiliation{Center for Theoretical Astrophysics, Los Alamos National Laboratory, Los Alamos, NM 87545, USA}
\affiliation{Computational Physics and Methods, Los Alamos National Laboratory, Los Alamos, NM 87545, USA}

\author[0000-0002-8825-0893]{Brandon L. Barker}
\affiliation{Center for Theoretical Astrophysics, Los Alamos National Laboratory, Los Alamos, NM 87545, USA}
\affiliation{Computational Physics and Methods, Los Alamos National Laboratory, Los Alamos, NM 87545, USA}
 
\begin{abstract}

From tabulated nuclear and degenerate equations of state to photon and neutrino opacities, to nuclear reaction rates: tabulated data is ubiquitous in computational astrophysics. The dynamic range that must be covered by these tables typically spans many orders of magnitude. Here we present a novel strategy for accurately and performantly interpolating tabulated data that spans these large dynamic ranges. We demonstrate the efficacy of this strategy in tabulated lookups for nuclear and terrestrial equations of state. We show that this strategy is a faster \textit{drop-in} replacement for linear interpolation of logarithmic grids.

\end{abstract}

%% Keywords should appear after the \end{abstract} command. 
%% The AAS Journals now uses Unified Astronomy Thesaurus concepts:
%% https://astrothesaurus.org
%% You will be asked to selected these concepts during the submission process
%% but this old "keyword" functionality is maintained in case authors want
%% to include these concepts in their preprints.
%\keywords{Classical Novae (251) --- Ultraviolet astronomy(1736) --- History of astronomy(1868) --- Interdisciplinary astronomy(804)}

\section{Introduction} \label{sec:intro}

Complex and computationally difficult microphysical data is often tabulated for efficient lookup. Common examples include finite temperature nuclear equations of state, such as those used in neutron star merger and core-collapse supernova modeling \citep{stellarcollapsetables}, terrestrial equations of state suitable for moon or planet formation or planetary defense modeling \citep{sesame}, tabulated photon \citep{Rogers92Opal,Iglesias96Opal,Zhu21DustOpacity,Mills24GRRMHDBH} and neutrino opacities \citep{BurrowsNeutrinos,OConnorNuLib,fornax} for use in radiation transport, and tabulated nuclear reaction rates \citep{REACLIB} for use in reaction networks.

A common thread in these disparate problems is the need to span many orders of magnitude. For example, a finite temperature nuclear equation of state in the Stellar Collapse\footnote{\url{https://stellarcollapse.org/microphysics.html}} database typically spans densities between $10^4$ and $10^{12}$ g$/$cm$^3$ and temperatures between $10^{-2}$ and $10^{2}$ MeV. For this reason, logarithmic interpolation, where the independent and dependent variables undergo a transformation of the form
\begin{eqnarray}
    \label{eq:log:transform:x}
    x' &=& \log_{10}x\\
    \label{eq:log:transform:y}
    y' &=& \log_{10}y
\end{eqnarray}
is often invoked. Interpolation may then be performed on the transformed variables $x'$ and $y'$, rather than $x$ and $y$ via usual techniques such as linear or higher-order Lagrange interpolation.

Logarithms are transcendental functions \citep{townsend1915functions}, meaning they cannot be computed with a finite set of simple algebraic operations such as addition and multiplication. Rather they must be approximated up to machine precision via a series expansion \citep{cody1980software}. The logarithmic transformations \eqref{eq:log:transform:x} and \eqref{eq:log:transform:y} are therefore computationally expensive and can be a bottleneck in simulations.

The transformation invoked need not be exactly logarithmic, however. The important traits of the transformation are that \textbf{(a)} it has approximately logarithmic spacing, that is
\begin{eqnarray}
    \label{approx:log:spacing}
    x\frac{\partial x'}{\partial x} \approx k
\end{eqnarray}
for some constant $k$ and that \textbf{(b)} it is exactly invertible. Exact invertibility is required to effectively transform in and out of the space when performing interpolation. Relation \eqref{approx:log:spacing} is required so that the a uniform spacing in $x'$ captures the range of scales in $x$. 

\begin{figure*}[t!]
    \centering
    \includegraphics[width=0.95\linewidth]{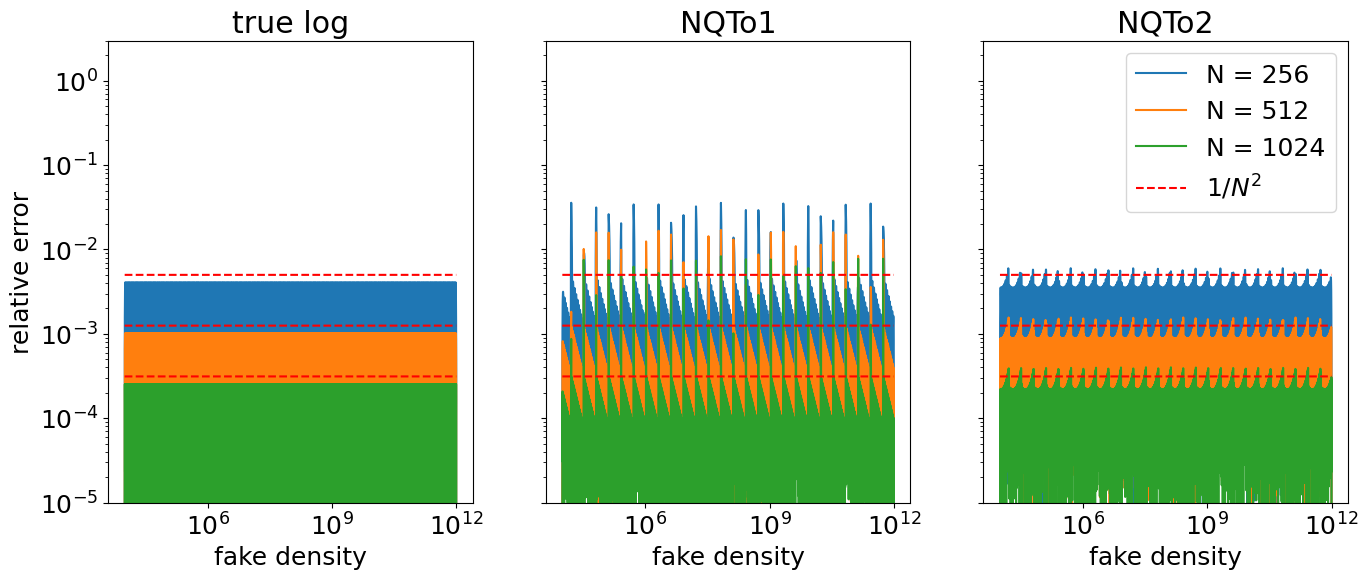}
    \caption{Linear interpolation of a synthetic function $P = 1 + \rho + K_1 \rho^\Gamma_1 + K_2 \rho^\Gamma_2$ on a true log grid (left), first-order NQT grid (center), and second-order NQT grid (right). The x-axis is $\rho$ and the y-axis is the relative error in $P$. On average (in the $L_1$ sense), all interpolants converge at second order, but error spikes near the control points of the NQT gridding converge more slowly. The result is that first-order NQT converges at second order in the $L_1$ norm but slower in the $L_p$ norm for $p>1$. In second-order NQT the spikes also converge at second-order, recovering second-order convergence in higher $p$ $L_p$ norms. In all three figures, a dashed horizontal line shows the expected magnitude for second-order convergence.}
    \label{fig:faketable}  
\end{figure*}

\citet{NQTo1} recognized that a transformation need not be logarithmic and introduced a family of exactly invertible functions that they called \textit{not-quite-transcendental} (NQT). These functions leverage the underlying structure of floating point numbers to construct highly performant almost-logarithmic functions. \citet{NQTo1} argued that these functions were drop-in replacements for normal logarithms in a wide variety of contexts, with no loss in accuracy. 

Unfortunately, the original NQT functions are exactly invertible, but they do introduce additional error in all $p-$norms for $p > 1$. For reasons that will be discussed below, we dub these original functions \textit{first-order} NQT functions, or NQTo1. Here we introduce a new family of NQT functions, which we dub \textit{second-order} NQT functions, or NQTo2. We show that second-order NQT functions provide significant improvement in accuracy over their first-order counterparts at acceptable increase in cost. We also demonstrate with several numerical experiments that they are excellent drop-in replacements for logarithms for second-order logarithmic interpolation of tabulated data. That is, if a code is currently performing linear interpolation on logarithmically spaced data, the NQTo2 method presented here will provide a speedup at essentially no cost in accuracy. However, higher-order interpolation, such as the biquintic interpolation used in the Helmholtz equation of state \citep{HelmEOS}, will not benefit.

This paper is organized as follows. In Section \ref{sec:NQTo1}, we review first-order NQT functions and describe why the original claim of \citet{NQTo1} that they imply zero loss in accuracy was incorrect. In Section \ref{sec:NQTo2}, we introduce second-order NQT functions and discuss why they cure the issues of the first-order version. In Section \ref{sec:numexperiments} we demonstrate the efficacy of NQTo2 methods in a suite of test problems. Finally in Section \ref{sec:outlook} we conclude our discussion. Appendix \ref{sec:integer:aliasing} discusses how to implement performant versions of the NQT method, which requires understanding the bit-level representation of floating point numbers.

\begin{figure*}[tb]
    \centering
    \includegraphics[width=0.99\textwidth]{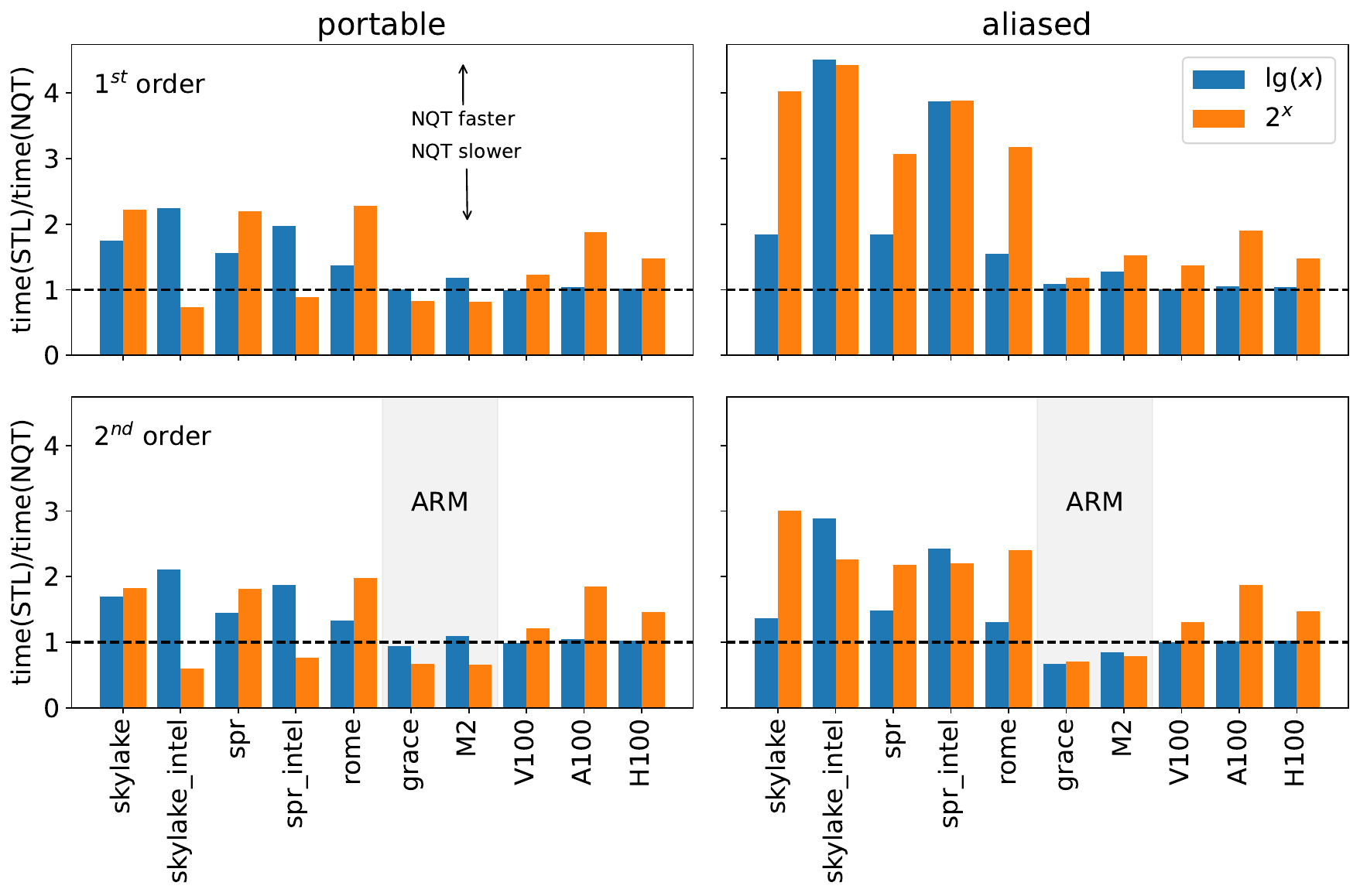}
    \caption{Ratio of performance of standard library provided functions to their NQT counterparts. Bigger is better. Left column is the portable implementation that uses {\tt frexp} and {\tt ldexp} and right column is the integer aliased implementation described in Appendix \ref{sec:integer:aliasing}. Top row is first-order NQT and bottom row is second-order NQT. In the architecture lists, {\tt \_intel} implies the intel compiler was used. On CPU systems, the gnu compiler was otherwise used and on GPU systems the relevant vendor-provided compiler (here CUDA) was used. {\tt spr} stands for Sapphire Rapids.}
    %\jf{This plot is hard to read at this scale.} \jmm{Made this a figure*}}
    \label{fig:NQT:comp}
\end{figure*}

\section{First-order not-quite-transcendental functions}
\label{sec:NQTo1}

A positive floating point number $n$ can be
represented as
\begin{equation}
    \label{eq:float:representation}
    n = m \times 2^{p},
\end{equation}
where $m\in [1/2, 1)$ is the \textit{mantissa} and the integer $p$ is
the \textit{exponent}. Most programming languages and hardware vendors
provide the ability to pull apart $n$ into its components, $m$ and
$p$.  This implies that
\begin{equation}
  \label{eq:approx:log}
  \lg(n) = \lg(m) + p
\end{equation}
which reduces the standard problem of computing a logarithm to
computing $\lg(m)$ on the interval $[1/2, 1)$. Change of basis
formulae can then move from $\lg$ to whatever logarithmic basis is
appropriate.

\citet{NQTo1} define  
\begin{equation}
  \label{eq:def:lg:NQT:o1}
  \lg_{o1}(n) = 2 (m - 1) + p,
\end{equation}
which is the piecewise linear interpolant of $\lg(x)$ with control points at powers of 2 \citep{Hall}.  As \citet{NQTo1} observe, however, it is also an independent function in its own right, and straightforwardly invertible by simply finding $m$ and $p$ and computing $m \times
2^p$. \citet{NQTo1} define this inverse function as $\text{pow2}_{o1}(x)$. Multiplication by a constant value is all that is required to convert these functions to NQT versions of $\log_{10}(x)$ and $10^x$.

Definition \eqref{eq:def:lg:NQT:o1} is everywhere continuous, but only piecewise $C^1$, as the derivative is discontinuous at the control points, powers of 2. When this transformation is applied to smooth data, the resulting function, which must then be interpolated, also becomes merely piecewise $C^1$. Unfortunately, second order convergence of linear interpolation is only guaranteed for everywhere $C^1$ functions.

The consequences of this non-smoothness are shown in the central panel of Figure \ref{fig:faketable}. We create a fake table of the form
$$P = 1 + \rho + K_1 \rho^\Gamma_1 + K_2 \rho^\Gamma_2$$
and interpolate it with linear interpolation on a log-log grid and on a NQTo1-NQTo1 grid. The former is shown on the left pane, the latter in the center. On average (in the $L_1$ sense), both interpolants converge at second order. However, error spikes near the control points of the NQT gridding converge more slowly. The result is that first-order NQT converges at order in the $L_1$ norm but slower in the $L_p$ norm for $p>1$.

\begin{figure*}[p!]
    \centering
    \includegraphics[width=0.75\linewidth]{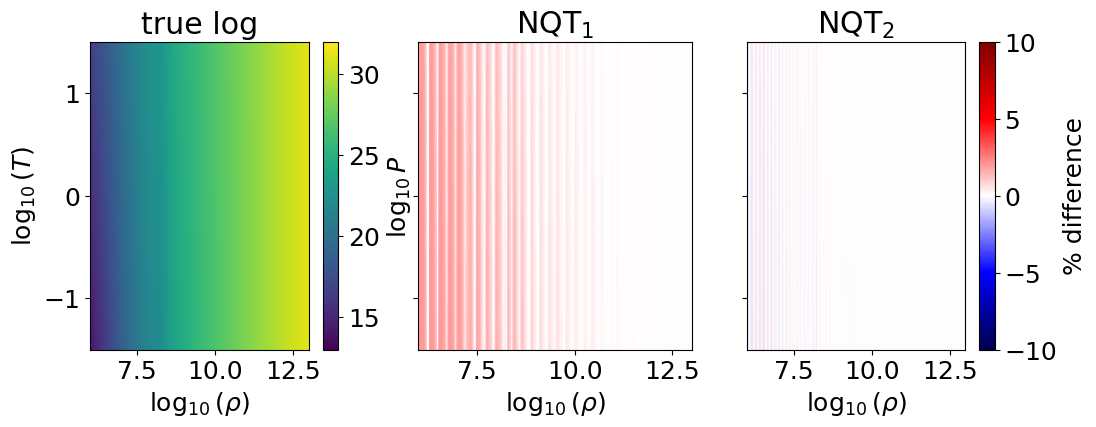}
    \caption{Interpolation of the SFHo EOS \citep{SFHoEOS} as tabulated in the Stellar Collapse database \citep{stellarcollapsetables}. Left column is the true table, middle is first-order NQT, and right is second-order. Left column shows pressure vs density and temperature at fixed $Y_e=0.1.$ Middle and right columns show relative error compared to the true log interpolant.}
    \label{fig:SFHo}
\end{figure*}

\begin{figure*}[p!]
    \centering
    \includegraphics[width=0.99\linewidth]{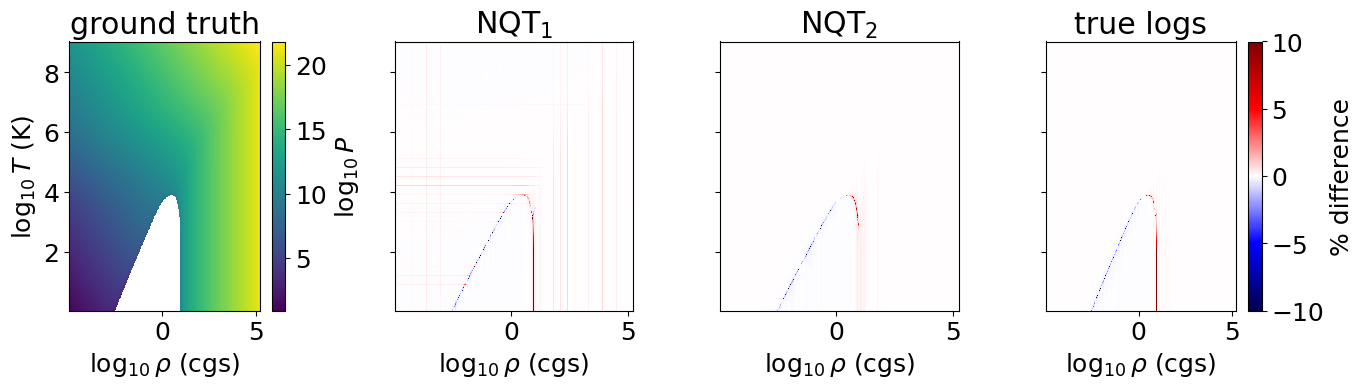}
    \caption{Interpolation of the tabulated equation of state for Copper in the Sesame database \citep{sesame,Peterson2012CopperEOS}. Left column is the fifth-order rational function interpolant produced by the EOSPAC library \citep{PimentelDavidA2021EUMV}, which we treat as ground truth. Middle two columns show errors from first- and second-order NQT interpolation respectively, while rightmost column shows error from true log interpolation.}
    \label{fig:copper}
\end{figure*}

\section{Second-order not-quite-transcendental functions}
\label{sec:NQTo2}

If not-quite transcendental functions can be made everywhere $C^1$, then second order convergence may be recovered. To do so, will modify definition \eqref{eq:def:lg:NQT:o1}. We begin with a second order polynomial form
\begin{equation}
    \label{eq:poly2}
    \lg_{o2}(x) = a x^2 + b x + c + p
\end{equation}
for some as-of-yet undetermined constants $a$, $b$, and $c$. To fix these, we enforce continuity of control points, that is
\begin{equation}
    \label{eq:control:cont}
    \lg_{o2}(2^p) = p
\end{equation}
and that
\begin{equation}
\label{eq:recurrence:gradients}
    \partial_x \lg_{o2}(x)\rvert_{x=2^p} = 2 \partial_x \lg_{o2}(x)\rvert_{2^{p+1}}
\end{equation}
which is indeed satisfied by a true logarithm.

This results in the \citet{Hermite1877} interpolant of the form
\begin{equation}
    \label{eq:def:lg:NQT:o2}
    \lg_{o2}(x) = - \frac{4}{3}*(m - 2)*(m - 1) + p.
\end{equation}
While intuitively four degrees of freedom (i.e., a cubic function) would be required to enforce continuity and continuity of a derivative, recurrence relation \eqref{eq:recurrence:gradients} allows the second order NQT function defined in \eqref{eq:def:lg:NQT:o2} to be everywhere $C^1$. And indeed, when one interpolates a function on an NQT mesh, second-order convergence is recovered, as shown in the right panel of \ref{fig:faketable}.

Equation \eqref{eq:def:lg:NQT:o2} is still exactly invertible, but the inverse now requires solving a quadratic equation for the mantissa:
\begin{equation}
    \label{eq:pow2:NQT:o2}
    m = \frac{3 - \sqrt{1 - 3*\lg_{o2}(m)}}{2}.
\end{equation}
While they are non-transcendental, numerically, square roots are also more computationally expensive than standard algebraic operations, limiting the utility of exponentiation.
%operation \jf{This is a nit-pick, but this isn't technically true. Square roots can be represented as the solution to a polynomial equation, so they're algebraic even though they can't be represented in a finite number of operations.}.
However, as we shall see in Section \ref{sec:numexperiments} performance is still superior to true logarithms and exponentials. 

Most programming languages provide facilities to split a floating point number into mantissa and exponent and to recombine. In C++ and Python these are called {\tt frexp} and {\tt ldexp} respectively. The NQT methods may be implemented in a portable manner via these functions. However, to achieve the best possible performance, the \textit{integer-aliasing} approach described in Appendix \ref{sec:integer:aliasing} must be used.

\section{Numerical Experiments}
\label{sec:numexperiments}

\subsection{Isolated performance}
\label{sec:experiment:performance}

To measure the performance of NQT functions compared to their transcendental counterparts, we implement performance portable versions of first- and second-order NQT logarithm and exponent in the Kokkos framework \citep{Kokkos}.\footnote{This implementation is available online at \url{https://github.com/lanl/not-quite-transcendental}.}

Figure \ref{fig:NQT:comp} compares timings for first- and second-order NQT methods for logarithm and exponential vs their transcendental counterparts across a variety of architectures including intel, AMD, and ARM CPUs as well as NVIDIA GPUs. Broadly, NQT methods are faster than transcendental methods, and the integer aliased methods are faster than the portable ones. Speedups for the second-order method are more modest than for first order. However, especially with the integer-aliased versions, 2x or greater speedups are achievable, even for the second-order NQT method.

Architecturally, speedups are more modest for GPU architectures than for x86-CPUs, likely because GPUs are memory bandwidth, not flop, limited. ARM64-based architectures sometimes indicate slowdowns rather than speedups (although the slowdowns are mild). This is likely due to the fact that ARM64 provides a hardware intrinsic for logarithms and exponentials, while x86 does not.
%This suggests that NQT functions are potentially only a win for x86 and GPU architectures, not ARM64. \jf{We should qualify this statement more carefully. AthenaK sees rather significant speedups on the Grace ARM CPU on Vista when using NQTs.} \jmm{Fair point. I'm fine just removing this statement.}

\begin{figure}[tb]
    \centering
    \includegraphics[width=0.9\columnwidth]{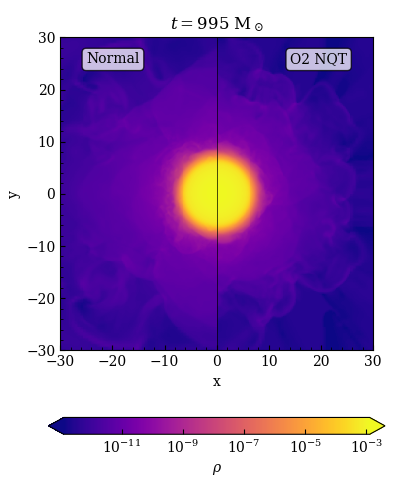}
    \caption{Density slice of a three-dimensional neutron star simulated with AthenaK with logarithmic interpolation on the left and second-order NQT on the right.}
    \label{fig:TOV:rho}
\end{figure}

\subsection{EOS lookups}
\label{sec:lookups}

To evaluate NQT methods integrated in table interpolation, we implement the integer-aliased versions of first- and second-order NQT methods inside the performance portable equation of state library {\tt Singularity-EOS} \citep{singularityeos}. Interpolation on device is efficated by the performance portable interpoaltion library, {\tt Spiner} \citep{Spiner}.

\begin{figure}[tb]
    \centering
    \includegraphics[width=0.99\columnwidth]{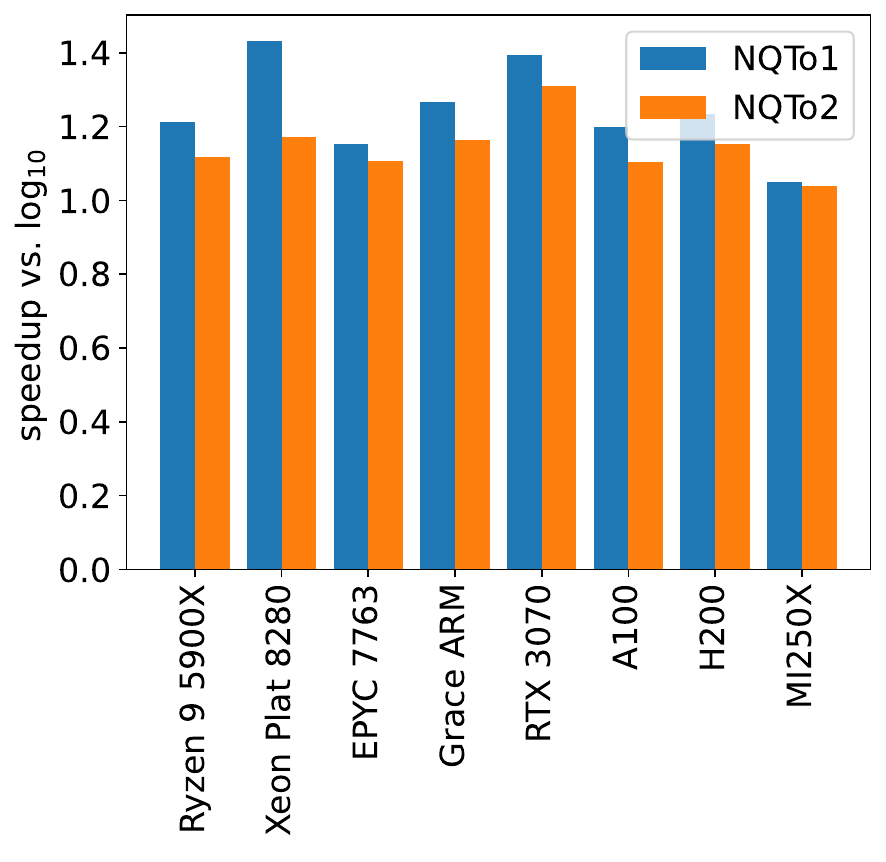}
    \caption{Integrated performance data comparing logarithmic with first- and second-order NQT interpolation in an AthenaK simulation of an isolated neutron star on various architectures. Speedups are with respect to logarithmic interpolation. Broadly NQTo2 always provides a speedup in integrated runtime, though smaller than NQTo1. The largest speedup for NQTo2 is about 30\% on an RTX 3070. Performance is measured by recording the total number of zone-cycles/wallsecond in each calculation.}
    \label{fig:TOV:perf}
\end{figure}

Figure \ref{fig:SFHo} shows interpolation of the SFHo \citep{SFHoEOS} equation of state as tabulated in the Stellar Collapse format \cite{stellarcollapsetables}. Particularly at low densities, the relative error for the second-order NQT is significantly reduced compared to first-order NQT, to the point that they are essentially indistinguishable. 

Figure \ref{fig:copper} compares interpolation treatments for the equation of state for Copper tabulated in the Sesame database \citep{sesame,Peterson2012CopperEOS}. The leftmost column pressure as a function of density and temperature computed via a fifth-order rational function interpolant computed by the EOSPAC library \citep{PimentelDavidA2021EUMV}, which we treat as ground truth. The middle two columns are the errors of first- and second-order NQT interpolation respectively, and the rightmost column is error from a true log grid. Copper undergoes a phase transition around standard density and temperature, where it goes into tension, interpreted as a negative pressure. This phase transition is sharp and none of the methods are correct, hence the cone-shaped error. This error cannot be removed. However, the logarithmic and second-order NQT interpolation are otherwise essentially indistinguishable, with very small errors. We note that these grids are not uniformly spaced in log space, rather they are more finely resolved around the phase transition.

In these experiments we find the integrated second-order NQT method to be about 2x faster than true logs on x86 architectures, and indistinguishable on GPU architectures, broadly consistent with the timing results reported in Section \ref{sec:experiment:performance}. In a pleasant surprise, we find the second-order NQT method to be about 25\% faster than true logs on ARM64. We are unsure why this is the case. However, given the artificial nature of the experiments in the previous section compared to these, perhaps a compiler optimization was available previously where now it is not.

\subsection{Integrated numerical relativity simulation}
\label{sec:experiment:integrated}

Finally as an integrated demonstration, we run simulations of an isolated neutron star, constructed using Tolman-Oppenheimer-Volkoff initial conditions \citep{Tolman1, Tolman2, TOV3} with a tabulated SFHo equation of state \citep{SFHoEOS}. These simulations are performed in the performance portable numerical relativity code AthenaK \citep{StoneAthenaKBase, Zhu24AthenaKNR, Fields24BNS}. A density slice is shown in Figure \ref{fig:TOV:rho}. Broadly results are similar. The eigenfrequencies of oscillations of an isolated neutron star depend sensitively on the equation of state, however, and we have not yet explored the density of table required for the NQTo2 and logarithmic tables to produce eiqenspectra that agree. 

\begin{figure}[t!]
    \begin{center}
    \resizebox{0.99\columnwidth}{!}{
    \begin{tikzpicture}
        \filldraw[draw=black,fill=blue,opacity=0.2] (0,0) rectangle ++(0.2, 1);
        \foreach \i in {1,...,11} {
            \filldraw[draw=black,fill=green,opacity=0.2] ({0.2*(0 + \i)},0) rectangle ++(0.2,1);
        }
        \foreach \i in {1,...,52} {
            \filldraw[draw=black,fill=red,opacity=0.2] ({0.2*(11 + \i)},0) rectangle ++(0.2,1);
        }
        \draw[<-,ultra thick] (0.1, 1) -- (0.1, 1.5) node[above] {\Large sign bit};

        \draw[|-|,ultra thick] (0.2, -0.25) -- ++({11*0.2}, 0);
        \draw ({0.2 + 11*0.1},-0.5) node[below] {\Large exponent}; 

        \draw[|-|,ultra thick] ({12*0.2}, 1.25) -- ++({52*0.2}, 0);
        \draw ({12*0.2 + 52*0.1},1.5) node[above] {\Large mantissa}; 
    \end{tikzpicture}
    }
    \end{center}
    \caption{Structure of a big-endian double precision floating point number under the IEEE754 standard.}
    \label{fig:bits}
\end{figure}
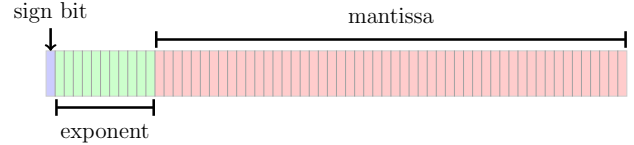

Performance data comparing logarithmic to first- and second-order NQT interpolation is shown in Figure \ref{fig:TOV:perf}. Improvements for second-order NQT are more modest than for first-order, but still are as large as approximately 30\%, a significant improvement. AthenaK makes many equation of state calls per cycle due to the conserved to primitive solver and first-order flux correction algorithm utilized, which may explain the large speedups.

\section{Outlook}
\label{sec:outlook}

To summarize, we have developed a novel set of functions that satisfy many of the properties that make logarithms appealing transformations for interpolation of tabulated data, but which are faster to evaluate, often significantly. We call these functions \textit{not-quite-transcendental}.

We have demonstrated in a series of experiments the efficacy of these NQT functions for the interpolation of tabulated equation of state data, such as finite temperature nuclear equations of state, suitable for core-collapse supernova and neutron star merger simulations, and terrestrial equations of state, suitable for planet and moon formation and planetary defense. 

This functionality has been implemented in the open source {\tt singularity-EOS} \citep{singularityeos} library and {\tt AthenaK} \citep{StoneAthenaKBase,Zhu21DustOpacity,Fields24BNS} and Phoebus \citep{Phoebus} numerical relativity codes, where it is now in use. While a portable implementation is possible, to squeeze the most performance out of the method, a lower-level, integer aliased approach is required. Reference implementations are available at \url{https://github.com/lanl/not-quite-transcendental}. 

Swapping out the logarithm can have a significant impact on performance. In AthenaK, we found that replacing the logarithm with NQTo2 in equation of state interpolation resulted in as high as a 30\% speedup in the total integrated calculation. This speaks to the large fraction of time that a numerical relativity simulation may spend in equation of state physics. There are still gains to be had in low-level optimizations such as this one.

\begin{acknowledgments}
The authors thank Luke Roberts, Josh Dolence, Patrick Mullen, Jeff Peterson, and Daniel Holladay for many useful discussions. PH and JMF acknowledge funding from the National Science Foundation under Grant No.~PHY-2116686. This research was supported by funding from the U.S. Department of Energy, Office of Science, Division of Nuclear Physics under Award Number DE-SC0021177. JMM was supported by the ASC M$^3$AP and XCAP projects in Advanced Simulation and Computing (ASC) program the at Los Alamos National Laboratory (LANL). 
BLB acknowledges support from the ASC Program of LANL as a Metropolis Postdoctoral Fellow.
LANL is operated by Triad National Security, LLC, for the National Nuclear Security Administration of U.S. Department of Energy (Contract No. 89233218CNA000001). This work is cleared for unlimited release under LA-UR-25-20182.
\end{acknowledgments}

\software{singularity-eos \citep{singularityeos}, 
          Spiner \citep{Spiner},
          Parthenon \citep{parthenon},
          Kokkos \citep{Kokkos},
          AthenaK \citep{StoneAthenaKBase,Zhu24AthenaKNR,Fields24BNS}, 
          Phoebus \citep{Phoebus},
          yt \citep{yt}
          }

%% Appendix material should be preceded with a single \appendix command.
%% There should be a \section command for each appendix. Mark appendix
%% subsections with the same markup you use in the main body of the paper.

%% Each Appendix (indicated with \section) will be lettered A, B, C, etc.
%% The equation counter will reset when it encounters the \appendix
%% command and will number appendix equations (A1), (A2), etc. The
%% Figure and Table counter will not reset.

\appendix

\section{Integer Aliasing}
\label{sec:integer:aliasing}

In this section, we discuss the \textit{integer aliasing} strategy required to implement the most performant versions of NQT functions.

\subsection{On the structure of floating point numbers}
\label{sec:floats}

The following discussion assumes a big-endian\footnote{That is to say the most significant bit is at the smallest memory address and the least significant bit at the largest.} implementation of the IEEE754 standard \citep{IEEE754-2019} for double precision numbers. However, it may be generalized to other precisions and other endianness straightforwardly.
Figure \ref{fig:bits} shows the structure of an IEEE754 number, which encodes equation \eqref{eq:float:representation}. Each box is a bit. The leading bit (at the lowest memory address) is the sign bit. The next 11 bits, $e$, are the exponent, represented as a biased integer in base 2, such as $p$ in equation \eqref{eq:float:representation}, where $p$ is given by $p=1-1023$. The remaining 52 bits represent the mantissa, again in base 2. The bits stored represent the number $0.b$, howver the IEEE754 calls for a leading 1, so the value is represented as $1.b$, which exists on the interval $[1,2)$. To recover $m\in [1/2, 1)$, we use $m=(1+b)/2)$. The mantissa and exponent may thus be extracted from the original number by appropriate masking of the bits. For example, to extract the mantissa $b$ of 64 bit floating point number $n$, one may use 
\begin{equation}
    \label{eq:mantissa:mask}
    m = (2^{52} - 1)\ \&\ n
\end{equation}
where $\&$ is the bitwise and operator. Similarly, the exponent $e$ may be extracted via
\begin{equation}
    \label{eq:exponent:mask}
    p = (2^{62} - 2^{52})\ \&\ n
\end{equation}
where equation \eqref{eq:mantissa:mask} selects for the trailing 52 bits and equation \eqref{eq:exponent:mask} selects for the exponent bits, but not the sign bit. For more details on constructing masks \eqref{eq:mantissa:mask} and \eqref{eq:exponent:mask}, see \citet{warren2012hacker}.

\subsection{Classic integer aliasing}
\label{sec:bithack:classic}

Because the leading order bits in the number represent the exponent, the total collection of bits reinterpreted as an integer scales roughly as the logarithm of the original floating point number. To be useful as a floating point number, the resultant integer must then be typecast back to its original type and then shifted and scaled, as shown below:
\begin{lstlisting}[caption={The basic integer-aliased NQTo1 log}, label={lst:nqto1}]
inline auto as_int(double f) {
  return *reinterpret_cast<std::int64_t*>(&f);
}

inline auto as_double(std::int64_t i) {
  return *reinterpret_cast<double*>(&i);
}

inline double lg_o1_aliased(const double x) {
  std::int64_t one_as_int = as_int(1.0);
  double scale_down = 1./static_cast<double>(as_int(2.0) - as_int(1.0));
  return static_cast<double>(as_int(x) - one_as_int) * scale_down;
}
\end{lstlisting}
This observation, made by \citet{Blinn}, is the core idea of \textit{integer aliasing}. Indeed it turns out the resultant function is exactly the first-order NQT $\lg$ defined in equation \eqref{eq:def:lg:NQT:o1}. Incidentally it is also the 64 bit generalization of the first part of the infamous Quake 3 fast inverse square root \citep{fastinvqrt,WizardryOfId}. The magic numbers from that algorithm can be mapped to the {\tt one\_as\_int} and {\tt scale\_down} numbers in Listing \ref{lst:nqto1}.

\subsection{Integer aliasing for second-order NQT methods}
\label{sec:aliasing:NQT:o2}

Integer aliasing for the second-order NQT method proceeds by picking out the mantissa and exponent as in equations \eqref{eq:mantissa:mask} and \eqref{eq:exponent:mask} and then using equation \eqref{eq:def:lg:NQT:o2}. To accelerate calculation, one may perform the arithmetic operations required, such as squaring the mantissa, while treating it as an integer, as shown below:
\begin{lstlisting}[caption={The integer-aliased NQTo2 log}, label={lst:nqto2}]
inline double lg_o2_aliased(const double x) {
  // picks out the mantissa
  constexpr std::int64_t mantissa_mask = (one << 52) - one;
  // picks out the least significant 26 bits
  constexpr std::int64_t low_mask = (one << 26) - 1;

  const std::int64_t x_as_int = as_int(x) - as_int(1);
  const std::int64_t m_as_int = x_as_int & mantissa_mask;
  const std::int64_t m_high = m_as_int >> 26;
  const std::int64_t m_low = m_as_int & low_mask;
  
  // square of the mantissa, but drops the least significant bits
  const std::int64_t m_squared =
      m_high * m_high + ((m_high * m_low) >> 25);

  return static_cast<double>(x_as_int + ((m_as_int - m_squared) / 3)) * scale_down;
}
\end{lstlisting}
Doing so will drop the least significant bits of the logarithm, however.

\bibliography{refs}{}
\bibliographystyle{aasjournal}

%% This command is needed to show the entire author+affiliation list when
%% the collaboration and author truncation commands are used.  It has to
%% go at the end of the manuscript.
%\allauthors

%% Include this line if you are using the \added, \replaced, \deleted
%% commands to see a summary list of all changes at the end of the article.
%\listofchanges

\end{document}